\documentstyle[aps,epsf]{revtex}

\begin{document}

\title{Quasiparticle Interactions  
in Two and Three Dimensional
Superconductors}

\author{D. Coffey}
\address{Department of Physics, State University of New York,
Buffalo, New York 14260}
\date{\today}

\maketitle
\widetext
\begin{abstract}
I investigate the difference between the quasiparticle properties
in two dimensional(2D)and three dimensional(3D) s-wave superconductors. 
Using the original BCS model for the pairing interaction and 
direct Coulomb interaction I show that 
quasiparticle interactions lead to a stronger energy dependence
in the single-particle self-energies
in 2D
than in 3D superconductors. This difference
arises from the presence of the low lying collective mode of the
order parameter in the 2D case which ensures that oscillator strength
in the response function is at low frequencies, $\sim \Delta$.
This strong quantitative difference between 2D and 3D
superconductors points to the importance of treating quasiparticle
interactions in low dimensional superconductors rather than assuming that
renormalizations remain unchanged from the normal state.
\end{abstract}

\newpage
Electronic properties in the superconducting
state have generally been investigated using mean-field
approximations.\cite{rev1,rev2}
In these calculations 
direct interactions
among quasiparticle in the ordered state
determined in the mean field approximation
are ignored.  These interactions
arise from the residual interactions neglected in the 
mean field 
approximation
but have been ignored with the assumption that 
any renormalization of quasiparticle properties determined in the normal state
remain unchanged in the superconducting state.
However quasiparticle interactions in the ordered state
have been shown to renormalize the superconducting gap,
change the temperature dependence of the gap,
and lead to finite quasiparticle 
lifetimes.\cite{bishop,coffey90,coffey93a,coffey93b,monthoux}
In the cuprate superconductors
quasiparticle interactions 
lead to a feature in tunneling conductance across an
superconductor-insulator-superconductor($g_{SIS}$) junction 
at $3\Delta$ in addition to
the peak at $2\Delta$ expected from the
mean field approximation.\cite{Zas,Hartge}  
The corresponding feature in the spectral density has also
been identified in ARPES data on BSCCO.\cite{Shen}

The question which I address in this paper is 
whether the 2D nature of the cuprate superconductors 
is responsible for the comparatively large effects of
quasiparticle interactions in the cuprates compared to 
conventional 3D superconductors.
In one dimension quantum fluctuations of the order parameter 
are sufficient to destroy long-range order\cite{Efetov}
 and in two-dimensions
thermal fluctuations destroy the order.\cite{Rice,Hohenberg}
Given this the effects of quantum fluctuations 
on superconducting properties
would be expected to
be more important in 2D than in 3D.
This is investigated 
using a model Hamiltonian 
for 2D and 3D s-wave superconductors at zero temperature
which includes the Coulomb interaction between electrons
and the pairing interaction originally introduced by BCS which
is characterized by a magnitude $V_0$ for electrons within an energy
$\omega_D$ of the Fermi surface.

The long-range nature of the Coulomb interaction
leads to a qualitative difference in the quasiparticle properties
of
2D and 3D s-wave superconductors.
The long-range nature of the
Coulomb interaction results in the
renormalization of the
Goldstone mode associated with the phase of the superconductor order
parameter in three dimensions 
so that the 
collective mode is at  the 
plasmon energy.\cite{PWA}
As a result the collective mode has no influence on superconducting
properties. In 2D
the collective mode remains gapless
just as the plasmon mode does for the 2D
electron gas in the normal state.
There is experimental evidence for low lying 
collective modes in the cuprate superconductors
and calculations also support the existence of low energy collective modes
in layered superconductors.\cite{Dunmore,Hwang}
This makes it possible for the collective mode
to have a strong influence on quasiparticle properties 
in the cuprates unlike the
case in conventional 3D superconductors.

Most models of the electronic properties of the cuprate
superconductors have emphasized their layered nature
and have employed 2D or
quasi-2D Hamiltonians.
These models have stressed short-range correlations
either through an on-site Hubbard repulsion, in the
weak or strong-coupling limits,
or phenomenologically through residual short-range antiferromagnetic
order which survives in the doped materials.
Except in references \cite{bishop,coffey90,coffey93a,coffey93b,monthoux} 
the effect of quasiparticle interactions
in the superconducting state have been ignored
and the importance of long-range correlations 
coming from the Coulomb interaction in quasi-2D systems,
such as the layered cuprate superconductors, have not been considered.

The single-particle self-energies are calculated here for 2D and 3D
s-wave superconductors 
and it is found that their magnitude in 2D is roughly ten times the value
in 3D.
The single-particle self-energies also have a stronger frequency 
dependence in 2D than in 3D
which leads stronger signatures of quasiparticle interactions
in the tunneling conductance of 2D superconductors compared to 
those in 3D superconductors.
 I briefly discussed these effects in 2D d-wave superconductors and in
quasi-2D superconductors.

The starting point of the present analysis is a Hamiltonian, Eq.(1),
describing a system of fermions interacting via Coulomb and pairing potentials,
\begin{equation}
H = \sum_{\vec k, \sigma} \xi_{\vec k} c^{\dagger}_{\vec k \sigma}
c_{\vec k \sigma}  
+
\sum_{\vec k\prime,\vec k,\vec q}
 V(\vec k, \vec q) c^{\dagger}_{\vec k - \vec q \uparrow}
c^{\dagger}_{-\vec k\prime + \vec q \downarrow}
     c_{-\vec k\prime \downarrow} c_{\vec k \uparrow}
+{{1}\over{2}}\sum_{\vec k\prime,\vec k,\vec q, \alpha, \beta}
U_{\vec q} c^{\dagger}_{\vec k - \vec q \alpha}
c^{\dagger}_{-\vec k\prime + \vec q \beta}
     c_{-\vec k\prime \beta} c_{\vec k \alpha}
\end{equation}
where $ \xi_{\vec p}= {{|\vec p|^2}\over{2m}}
-e_F$, $e_F={{ p_F^2}\over{2m}}$,
$p_F$ is the Fermi momentum,
 $U_{\vec q}={{2\pi e^2}\over{\epsilon_s |\vec q|}}$
for 2D and ${{4\pi e^2}\over{\epsilon_s |\vec q|^2}}$
for 3D, where $\epsilon_s$ is the dielectric constant of
the material,
and $V(\vec k, \vec q)$ is the same 
pairing interaction as introduced by BCS,
 $V(\vec k, \vec q)=-V_0\Theta(w_D-|\xi_{\vec k}|)
\Theta(w_D-|\xi_{\vec k -\vec q}|)$.
$w_D$ is a cutoff energy for the pairing interaction which is 
free parameter in the current calculation 
and was taken to be the 
Debye energy by BCS.
The Hamiltonian is written in terms of 
 quasiparticle operators,
$\hat \gamma_{\vec k \sigma}$,
which create elementary excitations of a mean field s-wave
superconducting ground state, 
by substituting $c_{\vec k \uparrow}$ with
$ u_{\vec k}\gamma_{\vec k \uparrow} 
+ v_{-\vec k}\gamma^{\dagger}_{-\vec k \downarrow}$
and $c_{\vec k \downarrow}$ with
$ u_{\vec k}\gamma_{\vec k \downarrow} 
- v_{-\vec k}\gamma^{\dagger}_{-\vec k \uparrow}$
 where $u^2_{\vec p}=1-v^2_{\vec p}
={{1}\over{2}}(1+{{\xi_{\vec p}}\over{E_{\vec p}}})$ are the usual
coherence factors, $E_{\vec p}=\sqrt{\xi^2_{\vec p}+\Delta^2}$ is
the quasiparticle spectrum and $\Delta$ is determined by the weak
coupling gap equation.
The transformed Hamiltonian has three vertices
$\gamma^{\dagger}_{\vec k_1+\vec q \sigma}
\gamma^{\dagger}_{-\vec k_1 -\sigma}
\gamma^{\dagger}_{-\vec k_2 \sigma}
\gamma_{\vec k_2+\vec q \sigma}$,
$\gamma^{\dagger}_{\vec k_1+\vec q \sigma}
\gamma^{\dagger}_{-\vec k_1 \sigma'}
\gamma_{-\vec k_2 \sigma'}
\gamma_{\vec k_2+\vec q \sigma}$ and
$\gamma_{\vec k_1+\vec q \uparrow}
\gamma_{-\vec k_1 \downarrow}
\gamma_{-\vec k_2 \downarrow}
\gamma_{\vec k_2+\vec q \uparrow}$
and their hermitian conjugates.
The matrix element associated with these involves 
combinations of the coherence factors which are different 
for the Coulomb and pairing interactions because of their different 
dependence on spin.
The single-particle self-energy for the 
$\gamma^{\dagger}_{\vec k_1 \sigma}
\gamma_{\vec k_1 \sigma}$
and $\gamma^{\dagger}_{\vec k_1 \uparrow}
\gamma^{\dagger}_{-\vec k_1 \downarrow}$
propagators
for 
superconducting quasiparticles
with the mean field spectrum, $E_{\vec k}$,
are calculated at zero temperature
in the generalized random phase approximation(GRPA).
%
In the GRPA
ring diagrams are summed in which the 
vertices are dressed by the pairing interaction.
This leads to an effective interaction given by\cite{DSarma}
\begin{equation}
V_e(\vec q, \omega)=
{{U_{\vec q}}\over
{1-U_{\vec q}\Pi(\vec q, \omega)}}
\end{equation}
where
\begin{equation}
\Pi(\vec q, \omega) = A_{00}(\vec q, \omega) -
{{V_0c_{20}^2(\vec q, \omega)}\over
{1+V_0B_{20}(\vec q, \omega)}},
\end{equation}
$A_{00}(\vec q, \omega)=\sum_{\vec k}
{{2m^2(\vec k,\vec q-\vec k)[ E_{\vec q- \vec k}+E_{\vec k}] }\over
{ \omega^2-(E_{\vec q- \vec k}+E_{\vec k})^2}}$,
$B_{20}(\vec q, \omega)=\sum_{\vec k}
{{l^2(\vec k, \vec q-\vec k)
[E_{\vec q- \vec k}+E_{\vec k}]}\over
{\omega^2-(E_{\vec q- \vec k}+E_{\vec k})^2}}$
and
$c_{20}(\vec q, \omega)=\sum_{\vec k}
{{(E_{\vec q- \vec k}+E_{\vec k})}\over
{E_{\vec q- \vec k}E_{\vec k}}}
{{\omega \Delta}\over
{\omega^2-(E_{\vec q- \vec k}+E_{\vec k})^2}}$.
$l(\vec p,\vec k)$, $m(\vec p,\vec k)$,  
and $n(\vec p,\vec k)$,  
are the usual combinations of coherence factors.\cite{JRS}
I first discuss the imaginary part 
of
the self-energies
for $E >0$
which  are
 given by
\begin{equation}
\Sigma''_{\gamma^{\dagger}\gamma}(\vec p,E)=
\sum_{\vec q} \Theta(x)
n^2(\vec p,\vec q-\vec p)
V''_e(\vec q, x),
\hskip 20pt
\Sigma''_{\gamma^{\dagger}\gamma^{\dagger}}(\vec p,E)=
\sum_{\vec q} \Theta(x)
m(\vec p, \vec p-\vec q)n(\vec p, \vec p-\vec q)
V''_e(\vec q, x)
\end{equation}
where $x= E-E_{\vec p -\vec q}$,
$\Theta(x)$
is zero for
$x < 0$ and is equal to one otherwise.
The imaginary part of the effective interaction is
\begin{equation}
V''_e(\vec q, \omega)= {{U^2_{\vec q}\Pi''(\vec q, \omega)}\over
{|1-U_{\vec q}\Pi(\vec q, \omega)|^2}}+
U_{\vec q}
\delta[1-U_{\vec q}\Pi(\vec q, \omega_{\vec q})]
\end{equation}
The second term gives the collective mode contribution.
Collective modes
have been investigated extensively for 3D
superconductors\cite{Martin}
 and more recently for 2D
and quasi-2D
superconductors\cite{DSarma,Hwang,cote,Griffin,vdMarel95a}.
As is well-known the energy of the collective mode,
$\omega_{\vec q}$, in 3D
superconductors is comparable to the plasma energy and as a result
the collective mode has no influence on properties at energies
$\sim \Delta$.
On the other hand 
$\omega_{\vec q} \sim \sqrt{q}$
at long wavelengths in 2D superconductors
and
$V_e(\vec q, \omega)$ has an imaginary part
given by $\pi Z_{\vec q}\delta(\omega -\omega_{\vec q})$
for values of $\omega \sim \Delta$.
Collective modes exhaust the $f-$sum rule 
at longwavelengths
and  $Z_{\vec q} 
\simeq n{{q^2}\over{2m}}{{1}\over{\omega_{\vec q}}}$.
However as $\omega_{\vec q} \rightarrow 2\Delta$
$Z_{\vec q}$ 
goes to zero 
and
the collective mode is
Landau damped in the continuum for 
$\omega_{\vec q} \geq 2\Delta$.
The collective excitation reemerges as a resonance above the continuum
at higher values of $|\vec q|$.
The result is that the effective interaction has strength at low energies,
$\sim \Delta$, in 2D in contrast to 3D.

The calculated imaginary part of 
$\Sigma_{\gamma^{\dagger}\gamma}(\vec p,E)$
is shown for a 2D s-wave superconductor
in Figure 1
for the parameter values $N(0)V_0=-.5$, $\omega_D=e_F/10$,
giving $\Delta=0.0273e_F$,
and
$\epsilon_s p_F = 25.12\AA^{-1}$.
In Figure 1 the dashed curve is the contribution due to scattering off
the
collective mode
and the solid
curve is the sum of this contribution
and the contribution
due to scattering from the continuum.
The collective mode contribution in 2D
has a threshold at $E \simeq E_{\vec p}$ since
the collective mode is confined to $|\vec q|'s \simeq 
p_F({{\Delta}\over{E_f}})^2$
 and leads to quasiparticle decay at energies below
the $3\Delta$ threshold for the continuum contribution.
Using the same parameters for the 3D case as in the 2D case, 
One finds that the energy dependence
of the continuum contribution
to $\Sigma''_{\gamma^{\dagger}\gamma}(\vec p,E)$
for $E > 3\Delta$ is determined by the increase in phase
space for decay.
The magnitudes of this contribution to the
self-energies in 2D are roughly an order magnitude
larger than in 3D pointing to the greater importance of
quantum fluctuations in 2D compared to 3D.
There is no contribution from the collective mode in 3D
so that the strong energy dependence  at $E \simeq 3\Delta$
is absent in 3D.
$\Sigma''_{\gamma\gamma}(\vec p,E)$
is very similar in magnitude and frequency dependence for $E_p$
and $E$ less than $3\Delta$ but is reduced in magnitude compared to
$\Sigma''_{\gamma^{\dagger}\gamma}(\vec p,E)$
at higher values of $E_p$ and $E$.
The real part of $\Sigma_{\gamma^{\dagger}\gamma}(\vec p,E)$
is\begin{eqnarray}
\Sigma'_{\gamma^{\dagger}\gamma}(\vec p,E)=
-{{1}\over{2}}\sum_{\vec q}\biggl[&n&^2(\vec p- \vec q, \vec p) \Biggl ( 
V'_e(\vec q, E-E_{\vec p -\vec q})
+\int^{\infty}_0{{d\omega}\over{\pi}}
{{2(E -E_{\vec p -\vec q})V''_e(\vec q, \omega)}
\over{(E -E_{\vec p -\vec q})^2-\omega^2 }}\Biggr )
\\ \nonumber
&-&m^2(\vec p- \vec q, \vec p)  
\Biggl (V'_e(\vec q, E+E_{-\vec p-\vec q}) 
 +\int^{\infty}_0{{d\omega}\over{\pi}}
{{2(E +E_{-\vec p -\vec q})V''_e(\vec q, \omega)}
\over{(E +E_{-\vec p -\vec q})^2-\omega^2 }}
\Biggr ) \biggr]
\end{eqnarray}
There is an analogous expression for 
$\Sigma'_{\gamma\gamma}(\vec p,E)$.
$\Sigma'_{\gamma^{\dagger}\gamma}(\vec p,E)$
is plotted in Figure 2 for a 2D superconductor.
The $E$ dependence of $\Sigma'_{\gamma^{\dagger}\gamma}(\vec p,E)$
for $E \leq 4\Delta$ is responsible for the "strong-coupling"
features in the calculated $g_{SIS}(eV)$ which will
be discussed below.
The frequency dependence of $\Sigma'_{\gamma \gamma}(\vec p,E)$
is almost exactly the same as in 
$\Sigma'_{\gamma^{\dagger} \gamma}(\vec p,E)$
except that the variation in magnitude
with frequency is about a factor of 8 smaller.
As in the case of the imaginary part of these self-energies
the magnitudes of the real parts
are roughly an order of magnitude greater 
in 2D compared to 3D and in contrast to the 2D case there is only
a very weak frequency dependence.

The same calculation can be carried out for a 2D d-wave
superconductor which arises in many models of 
superconductivity of the cuprates. 
The long range Coulomb interaction 
leads to quantitatively the same collective mode behavior\cite{Hwang}
and
so
of weight in the 
effective quasiparticle interactions at low frequencies is also
present for 2D d-wave superconductors.

Experimental
quantities which measure
the
convolution of two superconducting densities of
states such as the tunneling current across an SIS junction
 or optical conductivity of a superconductor
can be sensitive probes of
the effects of quasiparticle
interactions.
The tunneling current is calculated using the Hamiltonian introduced by
Cohen et al.\cite{Cohen},
$H_T=\sum_{R,L}T_{RL}\psi^{\dagger}_{R\sigma}\psi_{L,\sigma}+ h.c.$,
which describes the destruction of a normal state electron on the
left side of the junction and the creation of a normal state
electron on the right side, where $\sigma$ is the spin of the electron.
The tunneling matrix element, $T_{RL}$, is between the normal 
electron states on the two sides of the junction.
In the absence of spin-flip scattering in the barrier
the matrix element is spin independent and for tunneling between
s-wave conductors, discussed here, 
the matrix can be taken to be a constant.
In anisotropic cases in which there are strong bandstructure effects
and also depending on the nature of the junction
the tunneling matrix element can be strongly
momentum dependent. This is clearly seen in the break 
junction experiments of Hartge et al.\cite{Hartge}
For a discussion of the modeling of the tunneling matrix
in the context of the cuprates see reference\cite{lcoffey97}
and references therein.
Assuming a constant tunneling matrix element,
the tunneling conductance 
across an SIS 
junction, $g_{SIS}={{\partial I(V)}\over{\partial V}}$,
is \begin{equation}
\propto
{{\partial }\over{\partial V}}
{\int^{eV}_{0}} 
d\epsilon \biggl[N_0(\epsilon)
N_0(eV-\epsilon)
+N_1(\epsilon)
N_1(eV-\epsilon)
\biggr]
\nonumber
\end{equation}
$N_0(\epsilon)= 
\sum_{\vec p}u^2_{\vec p}
A_{\gamma^{\dagger}\gamma}(\vec p,\epsilon)
+v^2_{\vec p}A_{\gamma^{\dagger}\gamma}(-\vec p,-\epsilon)
$,
$N_1(\epsilon)=
\sum_{\vec p}{{\Delta}\over{2E_{\vec p}}}
\biggl[A_{\gamma^{\dagger}\gamma^{\dagger}}(\vec p,\epsilon)
+A_{\gamma\gamma}(-\vec p,-\epsilon)\biggr]$, 
$A(p,\epsilon)$ are spectral densities,
and $V$ is the bias across the junction.
In the case of
 optical conductivity
there is also 
 a combination of coherence factors 
in the expression
and the bias would be replaced by the photon energy.
The density of states in a 2D superconductor, $N(\epsilon)$,
 and
$g_{SIS}$
between two 
2D superconductors is compared with 
the mean field approximation,  $N^{MF}(\epsilon)$
and $g^{MF}_{SIS}$, in Figure 3.
The negative shift in
$\Sigma'_{\gamma^{\dagger}\gamma}(\vec p,\epsilon )$
with increasing $E_{\vec p}$
leads to a piling up of states at $\epsilon \simeq \Delta$ in $N(\epsilon)$
compared to  $N^{MF}(\epsilon)$.
This leads to a peak in the SIS tunneling current
which is responsible for the form of $g_{SIS}$
at $eV \simeq 2\Delta$.
The step feature at $eV \simeq 4\Delta$ comes mostly
from the form 
of $\Sigma'_{\gamma^{\dagger}\gamma}(\vec p,\epsilon)$
at $\epsilon \simeq 3\Delta$.
If $\Sigma''_{\gamma^{\dagger}\gamma}(\vec p,\epsilon)$
had a step like dependence on $\epsilon$ a dip feature would be seen at 
$eV=4\Delta$ analogous to the dip at $eV =3 \Delta$ seen in 
the $d$-wave case\cite{coffey93a,coffey93b}.
This form for the self-energy relies on bandstructure 
and enhanced quasiparticle interactions at frequencies
$\sim \Delta$ compared to higher frequencies.
Antiferromagnetic spin fluctuations 
due to short-range Coulomb correlations
lead to this 
form for the self-energy.\cite{Monthoux93}

The results of this calculation show that interactions among quasiparticles
in low dimensional superconductors
are much stronger than in conventional 3-D superconductors
due to long range correlations from the Coulomb interaction.
Low lying collective modes are also present in layered materials in 
which the Coulomb interaction is present between layers.\cite{DSarma} 
This suggests that in quasi-2D layered superconductors such as the 
cuprates the same enhancement of
interactions among
 superconducting quasiparticles should occur.
This offers an explanation for 
the magnitude of the dip feature seen in $g_{SIS}$
on some of the cuprates\cite{Zas,Hartge,Shen}
and suggests that quasiparticle interactions
may also have important consequences in other
classes of low dimensional superconductors.
In particular in quasi-one dimensional systems the collective mode
is known to be linear in
the wavenumber\cite{Mooji}
 so that the low energy effective interaction should
be enhanced above that of the quasi-2D case discussed here.
This suggests that the effects of quasiparticles should be stronger
in the quasi-one dimensional organic superconductors than in the cuprates.

I wish to thank J.S. Kim for technical assistance.
This work was supported by the New York State Institute for
Superconductivity(NY SIS).
 
\twocolumn
\begin{figure}[htb]
\epsfxsize=70mm
\epsfysize=65mm
\epsffile{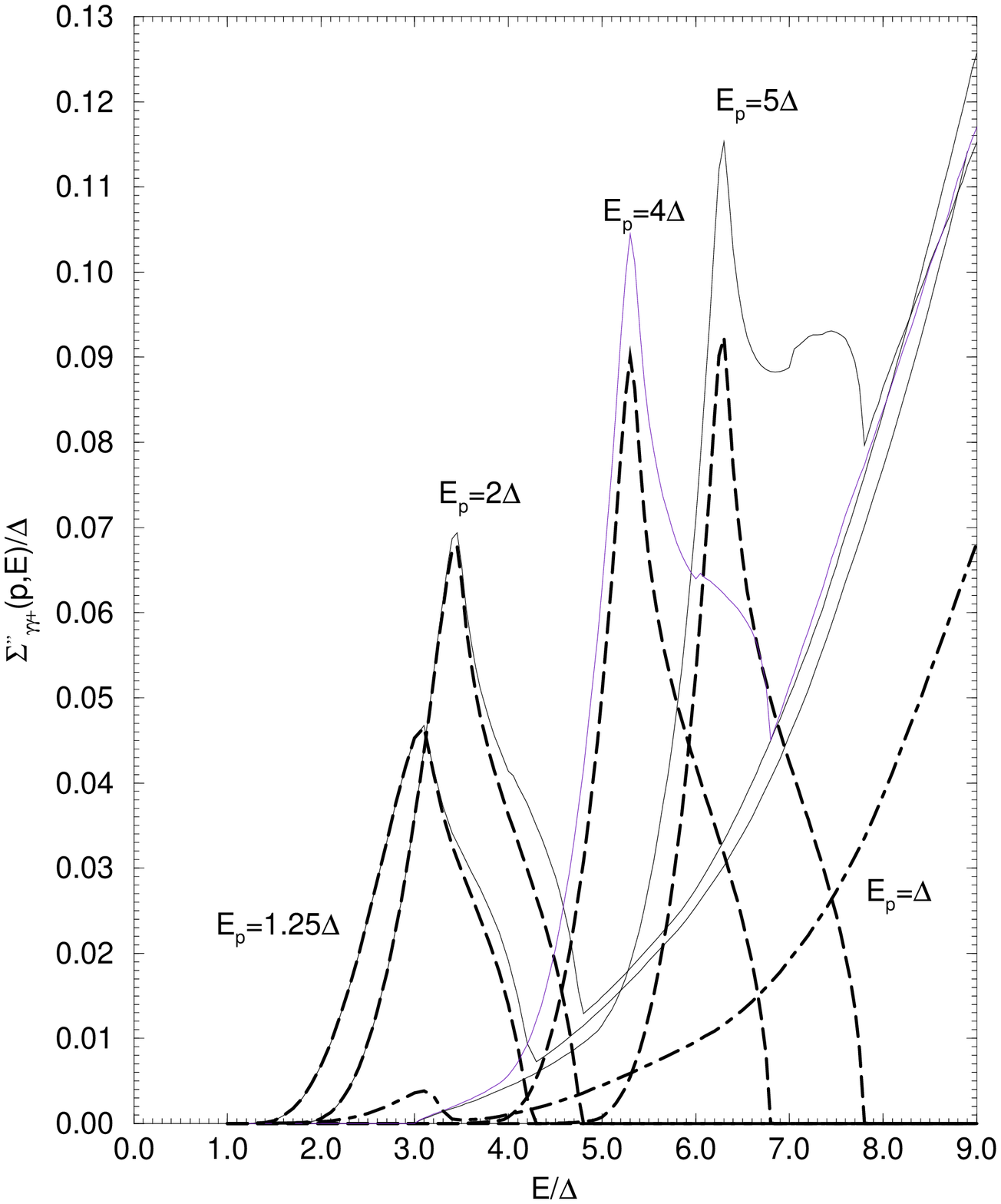}
\caption{Imaginary part of 
$\Sigma_{\gamma^{\dagger} \gamma}(\vec p,E)$
for a 2D s-wave superconductor for different values of $E_p$ with
$\epsilon_s p_F=25.13\AA^{-1}$, $\omega_D=0.1e_F$
and
$\Delta=0.0273e_F$.  
The dashed line is the contribution from the collective mode
for each $E_p$.
This is very small
for $E_{\vec p}=\Delta$(dot-dash curve).}
\label{fig1}
\end{figure}
\begin{figure}[htb]
\epsfxsize=70mm
\epsfysize=65mm
\epsffile{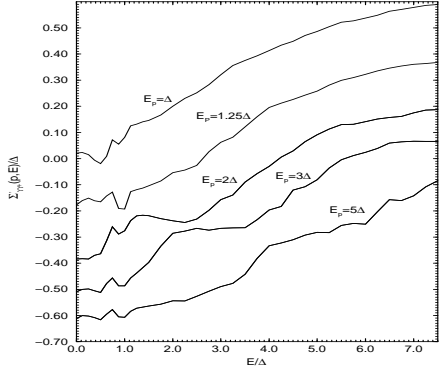}
\caption{Real part of the self-energy,
$\Sigma_{\gamma^{\dagger} \gamma}(\vec p,E)$,
for a 2D s-wave superconductor for different momenta with 
$\epsilon_s p_F=25.13\AA^{-1}$, $\omega_D=0.1e_F$
and
$\Delta=0.0273e_F$.}
\label{fig2}
\end{figure}
\begin{figure}[htb]
\epsfxsize=70mm
\epsfysize=65mm
\epsffile{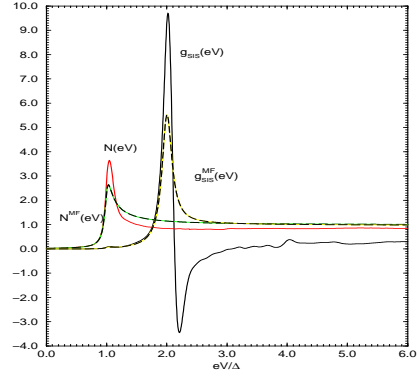}
\caption{Comparison between $N(\epsilon)$ and
$g_{SIS}$(solid curves)
with the mean field approximation, $N^{MF}(\epsilon)$
and $g^{MF}_{SIS}$(dashed curves),
for 2D superconductors.
$N^{MF}(\epsilon)$ was calculated using a lorentzian with
$\Gamma=\Delta/20$.}
\label{fig3}
\end{figure}

\end{document}